\documentclass[aps,prl,showpacs,twocolumn]{revtex4}
\usepackage{amssymb}
\usepackage{epsf}  %%% 

\begin{document}

\title{Torsional Strain of TaS$_3$ Whiskers on the Charge-Density Waves Depinning.}

\author{V.Ya. Pokrovskii, S.G. Zybtsev and I.G. Gorlova.}
\affiliation{Institute of Radioengineering and Electronics of RAS, 125009
Mokhovaya 11-7, Moscow, Russia.}

\begin{abstract}
We find that electric current $I$ exceeding the threshold value results in torsion deformation of o-TaS$_3$ samples with one contact freely suspended. 
The rotation angle, $\delta \phi$, of the free end achieves several degrees and exhibits hysteresis as a function of $I$.
The sign of $\delta \phi$ depends on the $I$ polarity; 
a polar axis along the conducting chains (the {\it c}-axis) is pointed out. 
We associate the effect with surface shear of the charge-density waves (CDW) coupled to the crystal shear. 
The current-induced torsional strain
could be treated in terms of enormous piezoelectric coefficients ($>10^{-4}$~cm/V) corresponding to shear. In essence, TaS$_3$ appears to be a ready torsional actuator based on the unique intrinsic property of the CDW.
\end{abstract}
\maketitle
Substantial progress in comprehension of the properties of the charge-density waves (CDW) in quasi one dimensional conductors has been achieved on considering the CDW as an electronic crystal developing inside the host lattice \cite{Grobzor}. 
Deformations and metastable states are intrinsic features of the CDW body \cite{Grobzor}. One can create uniform in average  (thermally-induced), or non-uniform along the sample ({\it e.g.} electric-field induced )
CDW deformations.
A more recent understanding is that the CDW deformation can result in the deformation of the crystal. Firstly the interaction was observed indirectly, as the softening of the lattice -- decrease of the Young modulus, $Y$, up to 4\%, \cite{BrillObzor,Mozurkth,Brill1,Mozurkexp,Brill2,PRL} and of the shear modulus, $G$, up to 30\% \cite{BrillObzor,BrillShear1,BrillShear2,Tas42Ishear} on the CDW depinning. 
According to \cite{Mozurkth}, the lattice and the CDW could interact as two springs connected somehow.
If so, pinned CDW contribute to the sample elastic modulus. Once depinned, the CDW rapidly relax, and their elastic contribution drops out \cite{Mozurkth}. 

Imagine another situation: the CDW are deformed by some external force. Due to the interaction of the two ``springs'' the crystal will change its dimensions, to enable the CDW to approach their equilibrium. 
This was experimentally observed as metastable length states
resulting from application of electric field \cite{HBZI} or thermocycling \cite{PRL}. Thus, length change on the CDW strain or compression can be directly put into correspondence  with the drop of $Y$ on CDW depinning. By analogy, what kind of effect could correspond with the $G$ drop? 
The latter has been studied by the method of torsional oscillations around the chain direction \cite{BrillShear1,BrillShear2,Tas42Ishear}.
So, the first expectation is to observe torsional electric-field induced deformation. 
However, torsional strain corresponds with {\it non-uniform} shear deformation in {\it different} plains $\|$ to the chains, its value growing proportionally to the distance from the rotation axis, whereas uniform shear would correspond to a change of the inter-face angle (parallelogram-type distortion) in some plain $\|$ to the chains. Further, one should find a mechanism transforming the axial (in-chain) field or CDW deformation into rotating force and a reason for the clockwise -- counter-clockwise asymmetry.

In the present Letter we report torsional strain of needle-like TaS$_3$ samples in the Peierls state on application of electric field above the threshold, $E>E_t$, and suggest answers to the questions above. 
The torsion deformation appears to be coupled with non-uniform CDW deformation;
it  reveals surface pinning and development of a polar axis along the chains direction -- the $c$-axis, likely due to a ferroelectric/ferroelastic-type transition. Up to our knowledge, we demonstrate the first torsional actuator, in which the torque is the intrinsic property of the working element: it is not achieved by special configuration of elements (\cite{Korea} and Refs. therein) or an external force (\cite{Zettl} and Refs. therein).

For the experiment we selected TaS$_3$ samples from a high-quality batch -- $E_t \lesssim 0.3$~V/cm. This compound showing CDW transition at $T_P=220$~K is widely studied \cite{Grobzor} and is known to exhibit  pronounced effects of the CDW on the host lattice \cite{BrillObzor,PRL}. We selected samples with the cross-section areas in the range 5--300~$\mu$m$^2$ and prepared contacts separated by about 3~mm. The sample surfaces under both contacts were covered with gold laser-evaporated in vacuum. 
One of the contacts was made of indium with the usual cold-soldering technique and fixed at the substrate, so that the other end of the sample was suspended above the substrate (Fig.~1). The contact to the hanging end was provided by a long  thin (typically 10~\textsf{x}~0.2~$\mu$m$^2$) wire soldered with a conducting epoxy. As the wires, thin whiskers of the high-$T_c$ superconductor Ba$_2$Sr$_2$CaCu$_2$O$_x$ (BSCCO) or NbSe$_3$ were used covered, again, with golden film.

\begin{figure} 
\epsfxsize=7cm
\leavevmode{\epsfbox{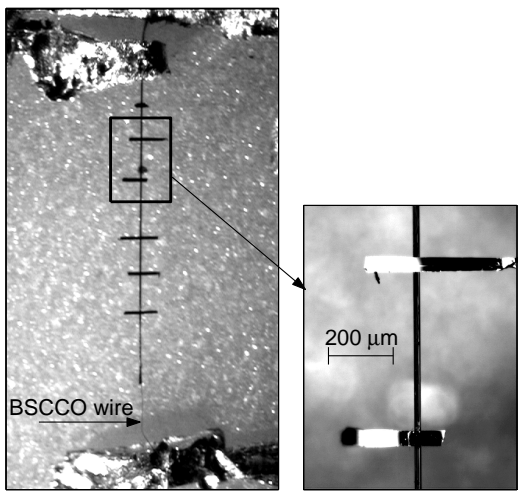}}
\caption{Left: A microphotograph of a TaS$_3$ sample with a suspended contact and 6 micro-mirrors attached. The contact separation is 3.4~mm. Right: An enhanced fragment of the same sample.}
%\label{expl}
\end{figure}

The configuration described allows nearly free torsional strain of the samples. We studied the deformation tracing the deflection of the laser beam reflected from the sample (inset to Fig.~1). With this purpose one or several micro-mirrors, made, again, from laser-cut BSCCO whiskers with golden films, were sticked to the samples. In some cases the reflection could be observed also directly from the sample surfaces. Three different techniques were used to trace the reflection displacement \cite{tbp}. The samples were placed in the optical cryostat; the measurements were performed near the liquid-nitrogen temperature, the lowest temperatures were achieved by means of the nitrogen pump-out. The heating by the laser was minimized down to $<1$~K to exclude effect on the rotation observed. Synchronously with the rotation angle $\phi$ the sample differential resistance $R_d$ was monitored with the conventional lock-in technique as a function of $I$.

For all of the 13 samples studied the rotation of the mirrors by the angle of the order of 1$^{\rm o}$ at $T=79 \div 84 $~K under application of electric field indicated the torsional deformation of the samples. The samples rotate around the $c$-axis, no substantial vertical shift of the reflection is seen. 
   
Fig.~2 shows typical $\delta \phi(I)$ and $R_d(I)$ dependences for two TaS$_3$ samples about 3~mm long. The mirrors are attached near the suspended contacts.  The arrows  show the direction of current sweep. 
One can see hysteresis loops in $\delta \phi(I)$ about 0.5$^{\rm o}$ wide. The changes of $\phi$ begin when the voltage is approaching the threshold. The most abrupt changes of $\phi(I)$ occur at currents slightly exceeding the threshold, then the dependence $\delta \phi(I)$ saturates. For some of the samples  growth  $|\delta \phi|$ gradually grows also for $|I|>|I_t|$ (in the saturation region).  The metastability of the sample torsional state at zero current correlates with 
the torque ``shape memory'' of applied torsion force reported in \cite{BrillShear2} and argues that the $\phi$ variation, or at least the principal part of it, is coupled to the CDW deformation, rather than to their dynamics. Alternatively, the elastic hysteresis could indicate a ferroelastic-type transition \cite{FerroBSCCO}.
    
\begin{figure} 
\epsfxsize=8.6cm
\leavevmode{\epsfbox{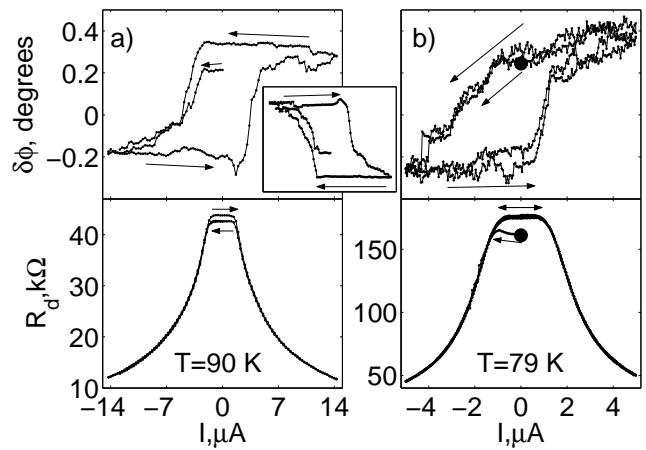}}
\caption{$\delta \phi(I)$ and $R_d(I)$ dependences measured simultaneously. a) and b) -- data for 2 different samples, the $\delta \phi$ scale is common. {\it Inset} to a) shows the $\delta \phi(I)$ curve for the sample from the same parent crystal with the $c$-axis turned over (arbitrary units). In b) the initial points ($I=0$) - the overcooled state - are marked with dark circles.}
%\label{expl}
\end{figure}

Note that the sign of $\delta \phi$ depends on the current direction. Thus, the torsional deformation cannot be attributed to some stress of the sample: according to \cite{BrillShear1,BrillShear2,BrillPrivat} $G$ drops for both current polarities, so, one could expect the same sign of $\delta \phi$ at positive and negative currents. Similarly, a Joule-heating origin of $\phi$ change rules out. We did not find a systematic correlation between the rotation direction and current polarity: 6 of 12 samples turned in one direction, and 6 -- in the opposite for the same current polarity. This could be expected: if we mentally turn a sample over,
we shall see that the same sign of $\delta \phi$ will happen for the opposite polarity. Other words, the effect could be observed only in the case of inequivalence of the sample ends. The randomness of the rotation direction indicates that this inequivalence cannot be attributed to the difference in the contacts type. 

An asymmetry of the sample ends is also revealed by the small loop on the $R_d(V)$ dependence (Fig.~2a) \cite{HBZI}. However, comparing the $\delta\phi(V)$ and $R_d(V)$ loops for different samples we did not find a correlation between their values and signs. Most likely, the inequivalence of the two current directions revealed by the $R_d(V)$ loop \cite{PRL,comment} does not concern the sign of $\delta\phi(V)$.

Another reason for the rotation observed could be a crystallographic asymmetry between the two directions along the $c$-axis. Orthorhombic TaS$_3$ belongs to the 222 point group \cite{Grobzor}, and in this case one can expect shear piezoelectric deformation in the $ab$ plain  if voltage is applied along $c$ \cite{Nai}.
However, the torsional strain observed is associated with the shear in the planes $\|$ to the $c$-axis. So, the deformation observed {\it cannot} be related to the symmetry group of the sample, if only TaS$_3$ does not undergo a ferroelectric transition...

To check the correlation of the rotation direction with the direction of the $c$-axis we cut a TaS$_3$ whisker perpendicular to the chains and turned one of the pieces by 180$^{\rm o}$, so that its $c$-axis appeared turned over. The two samples were positioned nearby on the same substrate, like the one in Fig.~1. The rotation directions appeared {\it different} for the same current polarity -- see Fig.~2a and the {\it inset} to it as an example. This result was reproduced for 4 pairs of whiskers. If TaS$_3$ has a polar axis, this could be expected in the context of the $c$-axis orientations.

We would like to emphasize that all the possible explanations having screw symmetry (a chiral axis \cite{c60}, a screw dislocation \cite{bronzeDisloc,GillScrew}) contradict the sign change of the torsional deformation on the sample  turn-over, which reveals a polar axis.

To examine the kind of internal force originating from the CDW deformations it would be important to study the torsion distribution along the samples. The multi-mirror configuration (Fig.~1) allows to do this. We found that all parts of the sample are turning in the same direction, the amplitude of $\delta \phi$ growing with the mirror distance, $x$, from the fixed probe. The dependences of the $\delta \phi$  swing ({\it i.e.} the width of the $\delta \phi(I)$ loop at $I=0$) on $x$  shown in Fig.~3, appears roughly linear. Similarly, shifting step by step the immobile contact towards the suspended one we found that the $\delta \phi$  swing falls approximately proportionally to the sample length. 

\begin{figure} 
\epsfxsize=5.4cm
\leavevmode{\epsfbox{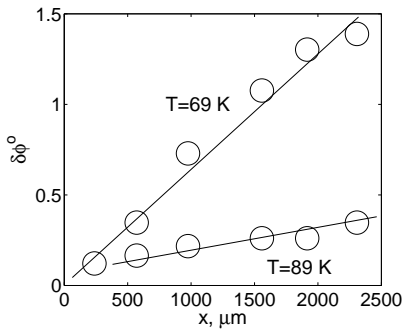}}
\caption{Dependences of $\delta \phi(I)$ amplitude on the distance from the fixed contact for 2 temperatures indicated in the plot. A sawtooth 0.1~Hz sweep of voltage with amplitude well above the threshold current was applied.}
%\label{expl}
\end{figure}

The dependence of $\delta \phi$ swing {\it vs.} $T$ appears rather strong: with $T$ decrease the span grows in average with the activation energy about 400~K -- half-width of the Peierls gap. This behavior is quite different from that of the length hysteresis loop \cite{PRL}, whose maximum is achieved around 100~K, and then drops gradually \cite{GolPokUnpub} (the measurements \cite{GolPokUnpub} were performed down to 35 K). The result indicates that the twisting can couple with the low-temperature anomalies \cite{Grobzor}, probably the lock-in, or/and the glass transition \cite{glass} (which might be related to the commensurability \cite{glass}): note, that the same activation energy, 400~K, is reported for the so-called $\beta$-process ($T<60$~K) \cite{glass}.

The $\delta \phi$ amplitude decreases with the growth of the sample width $w$ (Fig.~4). 
This could be expected: at fixed torque $\delta \phi \propto w^{-4}$, (see \cite{tork}, {\it e.g.}). In our case we can consider a rotating force acting uniformly either in a layer near the surface, or throughout the volume (the case of thin samples). In the 1st case the torque should be proportional to $w$, in the 2nd --- to $w^2$. Thus, one can expect $\delta\phi \propto w^{-3}$ or $w^{-2}$, which does not contradict the experiment (Fig.~4).

The frequency dependence of the $\delta \phi(f)$ amplitude needs a separate study. 
We noticed a slight decrease of  $\delta \phi$ with growing $f$ even for the slowest sweeps  ($f \sim 10^{-3}~Hz$). However, narrow resonance peaks $\delta \phi(f)$  for $f$ up to 2.4~kHz indicate a fast response of the sample. 

We also observe torsion of freely hanging end of the sample beyond the contact arising from the non-local nature of the CDW deformation.

\begin{figure} 
\epsfxsize=5.4cm
\leavevmode{\epsfbox{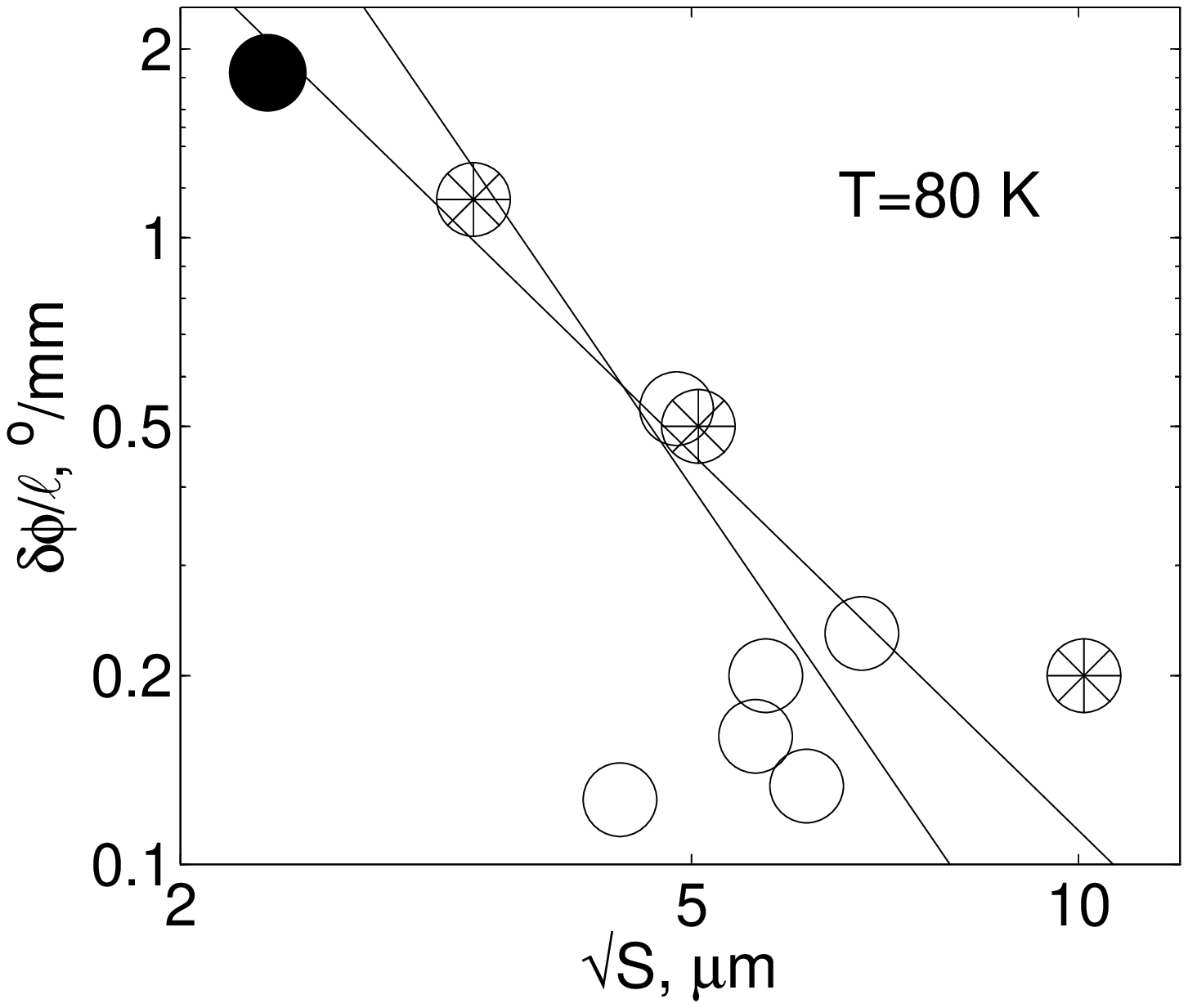}}
\caption{$\delta \phi$ amplitude normalized by the sample length $l$ {\it vs.}  $w \equiv \sqrt{S}= (\rho_{300}l/R_{300})^{1/2}$, where $\rho_{300}=3 \Omega \cdot \mu$m. $l=0.82$~mm for the thinnest sample (studied with another technique \protect\cite{tbp}) -- the dark circle, and $l\approx 3$~mm for the other samples. The circles marked with stars correspond to 3 different pieces of one parent sample. The solid lines show $\delta\phi \propto w^{-3}$ and $\delta\phi \propto w^{-2}.$}
%\label{expl}
\end{figure}

What kind of the CDW deformation gives rise to the torsional deformation? Clearly, electric field results in non-uniform CDW deformation along the sample, 
in contrast with uniform in average CDW deformation
achieved by thermal cycling.
This kind of deformation gives a much higher variation of resistance and other sample properties, including length \cite{PRL}. {\it E.g.}, the value of relative length change due to the uniform CDW deformation in TaS$_3$ achieves $\delta L/L = 5 \cdot 10^{-5}$ with $\delta R/R$ exceeding 30\%, while electric-field induced metastability gives $\delta L/L$ only $\sim10^{-6}$ with several percent $\delta R/R$ \cite{HBZI}. In contrast,
only a negligible hysteresis, within 0.1$^{\rm o}$, in $\delta \phi(T)$ was detected, while $\delta R/R$ achieved 30\% against the $<3$\% hysteresis in $\delta R/R$ {\it vs.} $I$. The same is illustrated by Fig.~2b. At $V=0$ (before starting the data acquisition) the current sweep was stopped, and the sample was heated by about 25~K by the laser and then cooled down. While the overcooled nature of the obtained state is obvious from the reduced value of $R$, no effect is seen on the $\delta \phi(I)$ dependence. Evidently, the observed torsional strain is somehow coupled to the non-uniform part of the CDW deformation. This could be the gradient of the CDW wave vector d$q$/d$x$ coupled with the contacts effects \cite{Grobzor}. However, as torsion strain implies high deformation of the crystal at the surface, it is more logical to relate the underlying CDW deformation to surface pinning, which can give rise to shear deformations of the CDW in the plains parallel to the $c$-axis \cite{qshear}. This assumption also agrees with the $\delta \phi$  distribution along the sample (Fig.~3) and its length dependence.
To our knowledge, it is the 1st observation of an effect of metastability related solely to non-uniform CDW deformation. 

The estimate $\delta \phi \frac{w}{l}$  gives shear at the sample surface $\sim10^{-4}$ and more. Dividing it by the electric field applied we find that the deformation is equivalent to that of a piezoelectric with the piezomoduli 
$d_{16}$ and $d_{15}$ exceeding $10^{-4}$~cm/V (index `1' corresponding to the $c$-axis direction).

What is the physical nature of such an enormous effect? A possible reason could be the variability of the S-S bond lengths with transferring electrons between the Ta and S atoms \cite{Meersho}. This feature of trichalcogenides is plausible for the explanation of coupling of the elastic properties \cite{BrillObzor} and deformations \cite{PRL,HBZI} with the longitudinal component of the $q$-vector. However, it can hardly concern the shear of the CDW. Alternatively, commensurability along the $c$-axis would provide direct coupling of the shear deformations of the CDW and the lattice. Surface pinning naturally gives shear in the plains $\|$ to the $c$ axis and $\perp$ to the crystal face. If a component of shear $\|$ to the face appears, it would result in torsion around the $c$-axis. This component can arise due to the symmetry loss of the TaS$_3$ lattice. We can assume that the low-$T$ anomalies -- a maximum of dielectric constant \cite{glass}, a lock-in transition, hysteresis in elastic properties -- can be the manifestations of a distributed \cite{APZZ} ferroelectric (ferroelastic) transition \cite{FerroBSCCO}.

In conclusion, we have observed electric-field induced torsional strain corresponding with enormous shear in whiskers of TaS$_3$. The threshold and hysteretic behavior of the torsion demonstrates its coupling with the CDW deformation. The sample deformation indicate surface pinning of the CDW and is induced only by non-uniform CDW deformation; in this sense it is a unique manifestation of metastability. The most puzzling result is indication of a polar axis in TaS$_3$. The strong $\delta \phi(T)$ dependence, different from that of $\delta L(T)$ \cite{PRL,GolPokUnpub}, could indicate a connection of the torsion effect with the glass \cite{glass} or/and the lock-in \cite{Grobzor} transition. Detailed structural studies of TaS$_3$ at low temperatures could be very promising. The particular kind of the CDW deformation inducing the torsion needs further clarification. The sample configuration (Fig.~1) appears to be a working torsional actuator, unique in the sense that the rotation moment arises from the intrinsic properties of the CDW, and does not require a special electric-field circuitry.

We are grateful to J.W. Brill, V.N. Timofeev, A.A. Sinchenko and S.V. Zaitsev-Zotov for useful discussions and to R.E. Thorne for the high-quality samples. The support of ISTC, RFBR (04-02-16509, 05-02-17578), RAS programs ``New materials and structures'' (No 4.21) and of the RAS Presidium is acknowledged. The research  was held in the framework of the CNRS-RAS-RFBR Associated European Laboratory ``Physical properties of coherent electronic states in condensed matter'' including CRTBT and IRE.

\end{document}